\documentstyle[11pt,epsf]{article}

\newcommand{\be}{\begin{equation}}
\newcommand{\ee}{\end{equation}}
\newcommand{\nn}{\nonumber}
\newcommand{\beba}{\begin{equation}\begin{array}{lcl}}
\newcommand{\eaee}{\end{array}\end{equation}}
\newcommand{\bea}{\begin{eqnarray}}
\newcommand{\eea}{\end{eqnarray}}
\newcommand{\ba}{\begin{array}}
\newcommand{\ea}{\end{array}}

\newcommand{\ns}{\normalsize}
\newcommand{\refs}[1]{(\ref{#1})}

\def\a{\alpha}
\def\b{\beta}
\def\g{\gamma}

\def\d{\delta}
\def\e{\epsilon}

\def\f{\phi}

\def\l{\lambda}
\def\m{\mu}
\def\n{\nu}

\def\r{\rho}
\def\s{\sigma}

\def\t{\tau}

\def\F{\Phi}

\def\O{\Omega}

\def\gh{\hat{g}}

\def\gh{\hat{g}}

\def\ght{\hat{g}\kern-0.6em \widetilde{\raisebox{-0.12em}{\phantom{X}}}}
\def\bht{\hat{b}\kern-0.6em \widetilde{\raisebox{0.15em}{\phantom{X}}}}

\def\cL{{\cal L}}
\def\cM{{\cal M}}
\def\cG{{\cal G}}

\begin{document}

%%%%%%%%%%%%%%%%%%%%%%%%%%%%%%%%%%%%%%%%%%%%%%%%%%%%%%%%%%%%%%%%%%%%%%%%%%%

\begin{titlepage}
\title{\hfill{\ns UPR-770T, HUB-EP-97/49\\}
       \hfill{\ns hep-th/9709030\\[.1cm]}
       \hfill{\ns August 1997}\\[.8cm]
       {\Large\bf U--duality Covariant M--theory Cosmology}}
\author{Andr\'e
        Lukas$^1\,$\setcounter{footnote}{0}\thanks{Supported by Deutsche
        Forschungsgemeinschaft (DFG) and
        Nato Collaborative Research Grant CRG.~940784.}~~and
        Burt A.~Ovrut$^1\, ^2\,$\setcounter{footnote}{3}\thanks{Supported in
        part by a Senior Alexander von Humboldt Award}\\[0.5cm]
        {\ns $^1$Department of Physics, University of Pennsylvania} \\
        {\ns Philadelphia, PA 19104--6396, USA}\\[0.3cm]
        {\ns $^2$Humboldt Universit\"at, Institut f\"ur Physik}\\
        {\ns Invalidenstra\ss{}e 110, 10115 Berlin, Germany}}
\date{}
\maketitle

\begin{abstract}
A manifestly U--duality covariant approach to M--theory cosmology is
developed and applied to cosmologies in dimensions $D=4,5$. Cosmological
properties such as expansion powers and Hubble parameters turn out to
be U--duality invariant in certain asymptotic regions. U--duality
transformations acting on cosmological solutions, on the other hand, shift
the transition time between two asymptotic regions and determine the details
of the transition. Moreover, in $D=5$, we show that U--duality can map
expanding negative and positive branch solutions into each other.
\end{abstract}

\thispagestyle{empty}
\end{titlepage}

%%%%%%%%%%%%%%%%%%

\section{Introduction}
With M--theory becoming the prime candidate for the underlying fundamental
theory, new phenomena in particle physics as well as early universe
cosmology have to be addressed. Consequently, there has been some
recent interest in M--theory cosmology. In a first paper~\cite{letter}
the authors have studied the r\^ole of form fields in type II and M--theory
cosmology. Subsequent papers~\cite{kal,lu,paper} gave broad classes of
cosmological solutions with form fields. In ref.~\cite{letter,paper} it
has been shown in detail that the main effect of these fields is to
connect Kaluza--Klein type rolling radii solutions~\cite{muell} with each
other. Various other aspects of M--theory cosmology have been studied so far,
such as the possibility of singularity free
solutions~\cite{paper,lw,ra,kfs,ko},
moduli and dilaton stabilization~\cite{dil} and the relation of
cosmological solutions and p--brane
solutions~\cite{letter,paper,ps,lw,lu1,lu2}.

The main focus of this paper is the relation of U--duality~\cite{Udual}
and cosmology, an aspect of M--theory cosmology which so far has not
received much attention. In the context of weakly coupled heterotic string
theory, T duality has been shown to be a useful tool in cosmology~\cite{Tcosm},
to generate solutions as well as to understand their
structure. Especially, the $O(d,d)$--covariant method of ref.~\cite{mv}
appears to be an extremely elegant approach to cosmology, which provides an
overview over a large class of solutions. At the same time it is closely
related to the development of pre--big--bang cosmologies~\cite{pbb}.
Applications of T--duality and S--duality to cosmology have also been
discussed in the ref.~\cite{ax_dil,li,ven,ba}. In ref.~\cite{ps,sen,clw}
the type IIB $SL(2)$ symmetry has been used to generate cosmological
solutions with Ramond--Ramond fields.

\vspace{0.4cm}

In this paper, we will present a manifestly U--duality covariant
formulation of M--theory cosmology. Since U--duality rotates metric degrees
of freedom and degrees of freedom from the 3--form of 11-dimensional
supergravity into each other, we will be naturally dealing with cosmologies
which have nontrivial ``Ramond--Ramond fields''. In particular, we will
recover some of the previously derived solutions and their
properties~\cite{letter,paper} in our new context.
The most interesting part of duality transformations in cosmology is the one
which acts non--trivially on the space--time metric and it is this part
on which we will concentrate. Therefore, we are going to reduce 11--dimensional
supergravity on a Ricci--flat manifold to $D$ dimensions, thereby keeping the
breathing mode of the internal space as the only modulus~\cite{maeda} and
focusing on the $D$--dimensional ``external'' part of U--duality. As a general
rule, this part of the U--duality group acting on cosmological solutions in
$D$ space--time dimensions is the same as the U--duality group of
$12-D$--dimensional maximal supergravity. As explicit examples, we will
study the cases $D=4,5$, corresponding to the U--duality groups
$G=SL(2)\times SL(3)$ and $G=SL(5)$, respectively. The example $D=5$ is
motivated by the Horava--Witten construction of M--theory on
$S^1/Z_2$ which represents the effective theory
of strongly coupled heterotic string theory. This theory turns out to be
effectively 5--dimensional for phenomenological values of the coupling
constants in some intermediate energy range. 

\vspace{0.4cm}

The outline of the paper is as follows. In the next section, we will perform
the reduction of 11--dimensional supergravity to $D$ dimensions, keeping
the ``minimal'' field content consistent with the external part of
U--duality. In section 3, we will work out the general U--duality covariant
formalism and present a general form for cosmological solutions. These
results will be applied to $D=4,5$ in the following two sections. We
conclude in section 6.

\section{Compactification of 11--dimensional Supergravity}

In this section, we are going to reduce the bosonic part of 11--dimensional
supergravity to $D$ space--time dimensions on a Ricci--flat manifold.
The resulting low energy action will be the starting point for our
discussion of cosmological solutions.

The bosonic part of the 11--dimensional supergravity Lagrangian reads
\be
 \cL = \sqrt{-\gh}\left[ \hat{R}-\frac{1}{2\cdot 4!}\hat{F}_{MNPQ}
       \hat{F}^{MNPQ}\right] +\frac{1}{3\cdot 3!(4!)^2}
       \,\e^{M_1...M_{11}}\hat{F}_{M_1..M_4}\hat{F}_{M_5..M_8}
       \hat{A}_{M_9M_{10}M_{11}}\; . \label{SUGRA11}
\ee
We are using the conventions of ref.~\cite{cj}. The 11--dimensional metric
and curvature are given by $\gh_{MN}$ and $\hat{R}$, respectively, where
uppercase letters are used to index the full space, that is,
$M,N,...=0,...,10$. The 4--form field strength $\hat{F}_{MNPQ}$ is expressed
in terms of the 3--form gauge field $\hat{A}_{NPQ}$ as
$\hat{F}_{MNPQ}=4\,\partial_{[M}\hat{A}_{NPQ]}$. For the class of
compactifications we will be interested in, the Chern--Simons term in
eq.~\refs{SUGRA11} vanishes. Therefore, we drop this term from now on and
consider the non--topological part of the Lagrangian
\be
 \cL = \sqrt{-\gh}\left[ \hat{R}-\frac{1}{2\cdot 4!}\hat{F}_{MNPQ}
       \hat{F}^{MNPQ}\right] \label{L}
\ee
only. Our main purpose is to investigate the relation of cosmological
solutions and U--duality symmetries for the action~\refs{L}. The focus in this
paper will be on the external part of U--duality which acts non--trivially
on the space--time metric rather than on the part which transforms moduli.
In our reduction to $D$ dimensions we will therefore keep a minimal moduli
content only, that is, the breathing mode of the Ricci--flat manifold.
Though formulae in this section are kept general, the most interesting cases
are the ones for $D=4,5$ on which we will concentrate later on. While the case
$D=4$ is of obvious relevance, the importance of $D=5$ is motivated by
the construction of Horava and Witten~\cite{hw} for the effective action
of the strongly coupled heterotic string. In this construction of
11-dimensional supergravity on the orbifold $S^1/Z_2$ times a 10--dimensional
manifold the orbifold direction turns out to be larger than the other
(six) compact directions for phenomenological values of the coupling
constants~\cite{w}. Consequently, the theory is effectively 5--dimensional
in some intermediate energy range.

\vspace{0.4cm}

Let us now be specific. We are using indices $\m ,\n ...=0,...,d\equiv
D-1$ for the external space--time, indices $m,n,...=1,...,d$ for the
external spatial directions and indices $a,b,...=d+1,...,10$ for the
$\d$--dimensional internal space, where $\d = 11-D$. Our Ansatz for the
11--dimensional fields is as follows
\bea
 \gh_{\m\n} &=& g_{\m\n}(x^\r ) \nn \\
 \gh_{\m b} &=& 0 \nn \\
 \gh_{ab} &=& \bar{b}^2(x^\r )\O_{ab}(x^c) \label{ans} \\
 \hat{A}_{\m\n\r} &=& B_{\m\n\r} (x^\s )\; . \nn
\eea
All other components of $\hat{A}_{NPQ}$ are set to zero. Here  
$\O_{ab}$ is the metric of a $\d$--dimensional Ricci--flat manifold and
$\bar{b}$ is its breathing mode. Depending on the dimension, this manifold
can be a Calabi--Yau space, a torus or even a product of both.
The $D$--dimensional metric and 3--form are denoted
by $g_{\m\n}$ and $B_{\m\n\r}$, respectively. As already discussed,
we have considered the minimal moduli content represented by $\bar{b}$ only.
As a further simplification, we have neglected $D$--dimensional
vector fields (graviphotons as well as those arising from $\hat{A}_{NPQ}$)
and $D$--dimensional 2 forms. In case of a reduction on a Calabi--Yau
manifold the latter are not present anyway, since the Betti number
vanishes $b_1=h^{1,0}+h^{0,1}=0$. With the above truncation, we arrive at a
low energy theory independent on the details of the compactification,
but we keep the characteristic 3--form as a low energy field. Furthermore,
as we will see, the Ansatz~\refs{ans} is consistent with the external part of
U--duality, so that it provides a ``minimal'' setting for our purpose.
Inclusion of other fields is straightforward but will not be considered
here since the minimal field content suffices to illustrate our main points.

Using the truncation~\refs{ans} the action~\refs{L} turns into
\be
 \cL = \sqrt{\O}\,\sqrt{-g}\;\bar{b}^\d\left[ R+
       \d (\d -1)\,\bar{b}^{-2}(\partial_\m\bar{b})^2-\frac{1}{2\cdot 4!}
       F_{\m\n\r\s}F^{\m\n\r\s}\right]\; , \label{L1}
\ee
where $F_{\m\n\r\s}=4\,\partial_{[\m}B_{\n\r\s ]}$. To get a canonical
curvature term we perform the Weyl rotation
\be
 g_{\m\n} = \bar{b}^{-\frac{2\d}{D-2}}\,\bar{g}_{\m\n} \label{Weyl}
\ee
to the Einstein frame metric $\bar{g}_{\m\n}$. In this frame, eq.~\refs{L1}
reads~\footnote{We drop the factor $\sqrt{\O}$ since it turns into a 
constant upon integration over the internal manifold.}
\be
 \cL = \sqrt{-\bar{g}}\left[ \bar{R}-k^2\,\bar{b}^{-2}(\partial_\m\bar{b})^2
       -\bar{b}^{\frac{6(11-D)}{D-2}}\frac{1}{2\cdot 4!}F_{\m\n\r\s}
       F^{\m\n\r\s}\right] \label{L2}
\ee
with $k^2=\frac{D-1}{D-2}\d^2-\d (\d -1)$. For a study of cosmological
solutions of this Lagrangian we consider the Ansatz
\bea
 \bar{g}_{\m\n} &=& \left(\ba{cc} -\bar{N}^2(\t ) & 0 \\
                                    0 & \bar{G}_{mn}(\t )\ea\right) \nn \\
 B_{mnr} &=& B_{mnr}(\t ) \\
 \bar{b} &=& \bar{b}(\t ) \nn \; .
\eea
Here, time has been denoted by $\t$.
The equations of motion with these specialized fields inserted can be
derived from a Lagrangian which is related to eq.~\refs{L2} by a formal
dimensional reduction to one dimension. This 1--dimensional Lagrangian
is given by
\bea
 \cL &=& \bar{N}^{-1}\sqrt{\bar{\F}}\left[ k^2\,\bar{b}^{-2}{\dot{\bar{b}}}^2
       -\frac{1}{4}{\bar{\F}}^2 {\dot{\bar{\F}}}^2-
       \frac{1}{4}\dot{\bar{G}}_{mn}{\dot{\bar{G}}}^{mn}\right. \nn \\
      && \quad\quad\quad\quad\quad
         \left. +\bar{b}^{\frac{6(11-d)}{D-2}}\bar{G}^{mm'}\bar{G}^{nn'}
         \bar{G}^{rr'}\dot{B}_{mnr}\dot{B}_{m'n'r'}\right] \label{L3}
\eea
where $\bar{\F}={\rm det}(\bar{G})$. The dot denotes the derivative with
respect to the time $\t$. 

In the last step we have dimensionally reduced $d=D-1$ dimensions of a theory
which by itself has been obtained reducing 11--dimensional
supergravity. Therefore, one should expect the U--duality group of
$(11-d)$--dimensional maximal supergravity as a symmetry group of the
Lagrangian~\refs{L3}. For example, for $D=4$ ($d=3$) the expected group is
the one of $8$--dimensional supergravity, that is, $G=SL(2)\times SL(3)$.
For $D=5$ ($d=4$) one expects $G=SL(5)$, the U--duality group of
7--dimensional supergravity. As we will show, this is indeed the case.
It is, however, hard to see directly from the Lagrangian in the form~\refs{L3}.
The reason is that we have performed a Weyl rotation~\refs{Weyl}
which is different from the one that leads to $(11-d)$--dimensional
supergravity with a canonical Einstein term. We can compensate for this
by the following nonlinear field redefinitions
\bea
 \bar{b} &=& \F^{-\frac{1}{2(10-D)}}\, b \nn\\
 \bar{N} &=& \F^{-\frac{9}{2(10-D)(D-2)}}\, b^{\frac{11-d}{D-2}}\, N \nn \\
 \bar{G}_{mn} &=& \F^{-\frac{11-d}{(10-D)(D-2)}}\,
                  b^{\frac{2(11-D)}{D-2}}\, G_{mn} \label{trafo}\\
 \bar{\F} &=& \F^{-\frac{(11-D)(D-1)}{(10-D)(D-2)}+1}\, b^{\frac{2(11-D)(D-1)}
              {D-2}} \nn\; ,
\eea
which express the physical Einstein frame fields $\bar{G}_{mn}$,$\bar{N}$,
$\bar{b}$, $\bar{\F}$ in terms of the new fields $G_{mn}$, $N$, $b$, $\F$
with 
\be
 \F = {\rm det}(G)\; .
\ee
Written in terms of these new variables the Lagrangian~\refs{L3} finally
reads
\bea
 \cL &=& N^{-1}\, b^\d\left[ -\d (\d -1)\,b^{-2}{\dot{b}}^2+\frac{1}{4(10-D)}
       \F^{-2}{\dot{\F}}^2-\frac{1}{4}\dot{G}_{mn}{\dot{G}}^{mn}\right.\nn\\
      &&\quad\quad\quad\quad
        \left. +\frac{1}{2\cdot 3!}G^{mm'}G^{nn'}G^{rr'}\dot{B}_{mnr}
        \dot{B}_{m'n'r'}\right]\; . \label{Lf}
\eea
This is the form of $\cL$ we are going to use in our discussion of U--duality
and cosmological solutions. For a physical interpretation of solutions one
should, of course, transform back to the Einstein frame fields
via eq.~\refs{trafo}.

\section{U--duality Covariant Formulation}

In this section, we will find the manifestly U--duality invariant form of
the Lagrangian~\refs{Lf} for $D=4,5$ and discuss its general solution.
Since the explicit parameterizations depend on the specific
U--duality group, we discuss the two cases $D=4,5$ separately.

\vspace{0.4cm}

$\underline{D=4}$~: The expected U--duality group in this case is
$G=SL(2)\times SL(3)$. Indeed, if we write the 3--form field $B_{mnr}$ as
\be
 B_{mnr} = \frac{B}{\F}\e_{mnr}\; ,\label{B_def}
\ee
with a scalar $B=B(\t )$, we can define $SL(2)/SO(2)$-- and
$SL(3)/SO(3)$--coset parameterizations $\cM$, $\cG$ by~\cite{memU}
\be
\cM = \F^{1/2}\left(\ba{cc} 1+\frac{B^2}{\F}&\frac{B}{\F}\\
                            \frac{B}{\F} & \frac{1}{\F}\ea\right)\; ,
      \quad\quad
\cG = \F^{-1/3}\left( G_{mn}\right)\; . \label{cos4}
\ee
Using these parameterizations and setting the internal dimension to $\d =7$,
Lagrangian~\refs{Lf} can be written as
\be
 \cL = N^{-1}b^7\left[ -42\,b^{-2}{\dot{b}}^2-\frac{1}{4}{\rm tr}
       \left(\dot{\cM}{\dot{\cM}}^{-1}\right) -\frac{1}{4}{\rm tr}
       \left(\dot{\cG}{\dot{\cG}}^{-1}\right)\right]\; . \label{L4}
\ee
In this form it is manifestly invariant under the $SL(2)\times SL(3)$
transformation
\bea
 b &\rightarrow& b \nn \\
 \cM &\rightarrow&P_2\,\cM P_2^T \label{trafo4}\\
 \cG &\rightarrow&P_3\,\cG P_3^T \nn
\eea
with $P_2\in SL(2)$, $P_3\in SL(3)$.

\vspace{0.4cm}

$\underline{D=5}$~: The U--duality group in this case is $G=SL(5)$. Let us
define the vector ${\bf B} = (B^s)$ by
\be
 B_{mnr} = \frac{1}{\F}\e_{mnrs}B^s\; .
\ee
Then the $SL(5)/SO(5)$ coset $\cM$ can be parameterized by~\cite{dl}
\be
 \cM = \F^{-2/5}\left(\ba{cc} G & -G{\bf B} \\
                              -{\bf B}^TG & \F+{\bf B}^TG{\bf B}\ea\right)\; ,
 \label{cos5}
\ee
where we have used a matrix notation $G=(G_{mn})$ for the metric. With
the internal dimension $\d =6$, Lagrangian~\refs{Lf} can then be put into
the form
\be
 \cL = N^{-1}b^6\left[ -30\, b^{-2}{\dot{b}}^2-\frac{1}{4}{\rm tr}\left(
       \dot{\cM}{\dot{\cM}}^{-1}\right)\right]\; ,
 \label{L5}
\ee
which has manifest $SL(5)$ invariance. The explicit transformations are
given by
\bea
 b &\rightarrow& b \nn \\
 \cM &\rightarrow& P\cM P^T
\eea
with $P\in SL(5)$.

\vspace{0.4cm}

Both cases $D=4,5$ can be treated uniformly by considering an
$SL(n)/SO(n)$ sigma model (where $n=2,3$ for $D=4$ and $n=5$ for $D=5$)
written in terms of the coset parameterization $M\in SL(n)/SO(n)$. It is
this model which we are going to solve in the remainder of this section.
Explicit applications to $D=4,5$ are the subject of the following sections.

Without reference to a specific parameterization, the coset $M$ can be
characterized by the conditions ${\rm det}(M)=1$ and $M=M^T$ which
can be implemented via Lagrange multipliers. We are therefore considering
the Lagrangian
\be
 \cL_1 = N^{-1}\, b^\d\left[ -\d (\d -1)\, b^{-2}{\dot{b}}^2+\frac{1}{4}
         {\rm tr}\left( M^{-1}\dot{M}M^{-1}\dot{M}\right)\right]
         +\l \left({\rm det}(M)-1\right)+{\rm tr}\left(\g (M-M^T)\right)
 \label{Lm}
\ee
with the Lagrange multipliers $\l$, $\g$. The $SL(n)$ symmetry transformations
are given by
\bea
 b \rightarrow b&,& \l \rightarrow \l \nn\\
 M \rightarrow PMP^T &,& \g \rightarrow {P^T}^{-1}\g P^{-1}\; ,
 \label{sln_trafo}
\eea
where $P\in SL(n)$. Infinitesimally, for $P\simeq 1+T$, ${\rm tr}(T)=0$ we
have
\be
 \d_T M = TM+MT^T\; ,\quad \d_T\g = -T^T\g -\g T \; .
\ee
The conserved current $J_T$ associated with an $SL(n)$ generator $T$ is
easily calculated to be
\be
 J_T = b^\d\,{\rm tr}\left( M^{-1}\dot{M}T\right)\; .
\ee
After eliminating the Lagrange multipliers, we find as the $SL(n)$
covariant equations of motion for $M$, $b$ and $N$ (in the gauge $N=1$ which
we can always choose by a suitable reparameterization of the time $\t$)
\bea
 \frac{d}{d\t}\left( M^{-1}\dot{M}\right) +\d HM^{-1}\dot{M} &=& 0 \nn \\
 (\d -1)\dot{H}+\frac{1}{2}\d (\d -1)H^2 &=& -\r \label{eoms}\\
 \frac{1}{2}\d (\d -1)H^2 &=& \r\; ,\nn
\eea
respectively. Clearly, the matrix $M$ in these equations is restricted to
be symmetric and unimodular. The Hubble constant $H$ and the energy density
$\r$ are defined by
\be
 H = \frac{\dot{b}}{b}\; ,\quad\quad \r = \frac{1}{8}{\rm tr}\left(
     M^{-1}\dot{M}M^{-1}\dot{M}\right)\; .\label{Hrho}
\ee
The second and third equation in~\refs{eoms} can be combined to find the
following solution for the breathing mode 
\be
 H = \frac{1}{\d \t}\; ,\quad\quad b=b_0\, |\t |^{1/\d}\; ,\label{b_sol}
\ee
where $b_0$ is an arbitrary constant. Inserting this into the first
equation~\refs{eoms} we find for the coset $M$
\be
 M = M_0\, e^{I\ln |\t |}\; ,\label{sol_M}
\ee
where the constant matrices $M_0$, $I$ satisfy
\bea
 {\rm det}(M_0) = 1&,&\quad {\rm tr}(I) = 0 \nn \\
 M_0=M_0^T &,&\quad M_0I=I^TM_0\; . \label{sol_cons}
\eea
Furthermore, from the last equation~\refs{eoms} one obtains the zero energy
constraint
\be
 {\rm tr}(I^2) = 4\frac{\d -1}{\d}\; .\label{E0}
\ee
Eqs.~\refs{b_sol}--\refs{E0} represent the complete solution of the
Lagrangian~\refs{Lm} written in a manifestly $SL(n)$ covariant form.
The $SL(n)$ transformation~\refs{sln_trafo} on the coset $M$ acts on the
integration constants encoded in $M_0$, $I$ by
\bea
 M_0&\rightarrow& PM_0P^T \nn \\
 I&\rightarrow& {P^T}^{-1}IP^T\; .\label{sln_par}
\eea
At this point, it is instructive to count the number of integration constants
in our general solution. The matrices $M_0$, $I$ satisfying the
constraints~\refs{sol_cons} contain $n^2+n-2$ independent parameters. The
zero energy condition~\refs{E0} eliminates one of them so that we remain with
$n^2+n-3$ independent integration constants. This is just about the correct
number to describe the general solution for all degrees of freedom in
the coset $M\in SL(n)/SO(n)$. On the other hand, the group $SL(n)$
consists of $n^2-1$ parameters which implies that for $n>2$ not all solutions
can be connected to each other by $SL(n)$ transformations. More precisely,
the total $n^2+n-3$--dimensional solution space splits into
$n^2-1$--dimensional equivalence classes, each consisting of solutions
related to each other by $SL(n)$ transformations via eq.~\refs{sln_par}.
The remaining $n-2$ integration constants label different equivalence classes,
that is, classes of solutions which cannot be connected by $SL(n)$
transformations. It is useful in the following to make this structure more
explicit in the solution~\refs{sol_M}. Diagonalizing $M_0$ and $I$ using
eq.~\refs{sln_par} with appropriate matrices $P$, it is straightforward to
prove that eqs.~\refs{sol_M}, \refs{sol_cons}, \refs{E0} can equivalently
be written in the form
\be
 M = P\, {\rm diag}(|\t |^{p_1},...,|\t |^{p_n})\, P^T
 \label{sol_M1}
\ee
with
\be
 \sum_{i=1}^{n}p_i = 0\; ,\quad\quad \sum_{i=1}^{n}p_i^2=4\frac{\d -1}{\d}
 \label{p_cons}
\ee
and $P\in SL(n)$. The equivalence classes of $SL(n)$ unrelated solutions
are parameterized by the $n-2$ constants $\{ p_i\}$ subject to the
constrains~\refs{p_cons}. In addition, since $SL(n)$ contains
permutations of the $n$ directions we should pick a definite order, say
$p_i\geq p_j$ if $i<j$, for the $\{ p_i \}$ to describe $SL(n)$ inequivalent
solution. On the other hand, a specific class, characterized
by a fixed set $\{ p_i\}$, is generated by the matrices $P\in SL(n)$ in
eq.~\refs{sol_M1}. In the next sections we will apply these general results
to the examples $D=4,5$ and discuss physical the implications in detail.

\section{The Example $D=4$}

We are now going to analyze the solutions of the 4--dimensional theory,
specified by the Lagrangian~\refs{L4}, in detail. The dimension of the internal
manifold is $\d = 7$. This manifold can, for example, be a torus $T^7$ or
a product of a Calabi--Yau 3--fold and $S^1$. The general equations of
motion~\refs{eoms}, \refs{Hrho}, adapted to the situation of two cosets
(with $\d =7$ inserted), now read
\bea
 \frac{d}{d\t}\left( \cM^{-1}\dot{\cM}\right)
          + 7\, H\cM^{-1}\dot{\cM} &=& 0 \nn \\
 \frac{d}{d\t}\left( \cG^{-1}\dot{\cG}\right)
          + 7\, H\cG^{-1}\dot{\cG} &=& 0 \nn \\
 6\,\dot{H}+21\, H^2 &=& -\r \\
 21\, H^2 &=& \r \nn
\eea
with
\be
 H=\frac{\dot{b}}{b}\; ,\quad \r = \frac{1}{8}{\rm tr}\left( \cM^{-1}
   \dot{\cM}\cM^{-1}\dot{\cM}\right) + \frac{1}{8}{\rm tr}\left( \cG^{-1}
   \dot{\cG}\cG^{-1}\dot{\cG}\right) \; .
\ee
The two matrices $\cM$, $\cG$ parameterize the cosets $SL(2)/SO(2)$,
$SL(3)/SO(3)$ respectively, and are both symmetric and unimodular. Their
explicit form in terms of the metric and the 3--form has been given in
eq.~\refs{cos4}. According to eq.~\refs{trafo}, the physical Einstein frame
fields can be expressed as
\bea
 \bar{b} &=& \F^{-1/12}b \nn \\
 \bar{N} &=& \F^{-3/8}b^{7/2} \label{phys4}\\
 \bar{G}_{mn} &=& \F^{-1/4}b^7\cG_{mn} \nn\; .
\eea
From eq.~\refs{b_sol}, the solution for the breathing mode reads
\be
 H = \frac{1}{7\t}\; ,\quad b = b_0\, |\t |^{1/7}\; .
\ee
For the cosets, we have from eq.~\refs{sol_M1}, \refs{p_cons}
\be
 \cM = P_2\,{\rm diag}(|\t |^p,|\t |^{-p})P_2^T\; ,\quad
 \cG = P_3\,{\rm diag}(|\t |^{p_1},|\t |^{p_2},|\t |^{p_3})P_3^T\; ,
 \label{sol_M4}
\ee
where
\be
 \sum_{i=1}^3p_i = 0 \; ,\quad 2p^2+\sum_{i=1}^3p_i^2 = \frac{24}{7}
 \label{p_cons4}
\ee
and $P_2\in SL(2)$, $P_3\in SL(3)$. As discussed before, we may require
$p>0$ and $p_1>p_2>p_3$. Totally, the solution is parameterized
by 13 independent integration constants. Two out of the four parameters
$p,p_1,p_2,p_3$ are independent, taking into account the
constraints~\refs{p_cons4}.
They describe the equivalence classes of $SL(2)\times SL(3)$--unrelated
solutions. The remaining 11 parameters in $P_2$, $P_3$ generate
$SL(2)\times SL(3)$--related solutions. As can be seen from eqs.~\refs{cos4},
\refs{trafo4} the $SL(3)$ subgroup is part of the global
coordinate transformations and does not generate independent new solutions.
Without loosing information, we can therefore set $P_3={\bf 1}$ in
eq.~\refs{sol_M4} and concentrate on the $SL(2)$ part of the U--duality
group. We write explicitly
\be
 P_2 = \left(\ba{cc} \a&\b\\ \g&\d\ea\right)\; ,\quad \a\d -\b\g = 1\; .
\ee
Then, from eq.~\refs{sol_M4} we get
\be
 \cM = \left(\ba{cc} \a^2|\t |^p+\b^2|\t |^{-p}&
                     \a\g |\t |^p+\b\d |\t |^{-p} \\
                     \a\g |\t |^p+\b\d |\t |^{-p}&
                     \g^2|\t |^p+\d^2|\t |^{-p}\ea\right)\; .
\ee
By comparison with the general form of the coset~\refs{cos4} we can read
off the fields $\F$, $B$. Recall that $B$ is the degree of freedom from
the 3--form defined in eq.~\refs{B_def}. To find the Einstein frame
fields $\bar{b}$, $\bar{N}$ and $\bar{G}_{mn}$ we insert the expression
for $\F$ into eq.~\refs{phys4}. The physical fields, written
in terms of the $SL(2)$ parameters, are then given by 
\bea
 \bar{b} &=& b_0\, (\g^2|\t |^p+\d^2|\t |^{-p})^{1/6}|\t |^{1/7} \nn \\
 \bar{N} &=& b_0^{7/2}(\g^2|\t |^p+\d^2|\t |^{-p})^{3/4}|\t |^{1/2} \nn\\
 \bar{G}_{mn} &=& b_0^7(\g^2|\t |^p+\d^2|\t |^{-p})^{1/2}|\t |^{p_m+1}\d_{mn}
 \label{solb4} \\
  B &=& \frac{\a\g |\t |^p+\b\d |\t |^{-p}}{\g^2|\t |^p+\d^2|\t |^{-p}}\; .\nn
\eea
Note that, though $\t$ was the comoving time in the original frame, this is
no longer true in the Einstein frame, since $\bar{N}\neq 1$. The comoving
time $t$ can be found by integrating the relation $dt = \bar{N}d\t$.
This can be explicitly carried out in two limiting cases and results in
\bea
 t &=& b_0^{7/2}|\g |^{3/2}\frac{2}{3+3p/2}|\t |^{3p/4+3/2}\,{\rm sgn}(\t )\; ,
 \quad  {\rm for}\quad |\t | \gg \t_{\rm form} \nn\\
 t &=& b_0^{7/2}|\d |^{3/2}\frac{2}{3-3p/2}|\t |^{3p/4-3/2}\,{\rm sgn}(\t )\; ,
 \quad {\rm for}\quad |\t | \ll \t_{\rm form}\; ,
 \label{asympt}
\eea
where
\be
 \t_{\rm form} = \left(\frac{\d}{\g}\right)^{2/p}\; .
\ee
The most relevant objects for the physical discussion are the scale factors
$a_m$ in the three spatial directions defined by $\bar{G}_{mn}=a_m^2(\t )
\d_{mn}$ and the corresponding Hubble parameters
$H_m\equiv\frac{1}{a_m}\frac{da_m}{dt}$. In the two limiting
cases~\refs{asympt}, the latter turn out to be
\be
 H_m =\frac{P_m}{t}\; ,\quad P_m = \left\{\ba{lllll}
      \frac{p_m+1-p/2}{3+3p/2}&&{\rm for}&&
      |\t |\gg\t_{\rm form}\\
      \frac{p_m+1+p/2}{3-3p/2}&&{\rm for}&&
      |\t |\ll\t_{\rm form}\ea\right.\; .
 \label{Hub4}
\ee
This shows, that in the asymptotic regions indicated in eq.~\refs{Hub4},
the expansion follows a power law with the powers determined by the
parameters $p,p_i$. The expansion (or contraction) rate is generally
different for the early and late asymptotic stage. Such behaviour has been
first described in ref.~\cite{letter,paper}. There, it has been shown, that
the asymptotic regions in \refs{asympt} correspond to Kaluza--Klein type
rolling radii solutions and the 3--form is effectively turned off in
these regions. The form is, however, operative at
\be
 |\t |\sim\t_{\rm form}=\left(\frac{\d}{\g}\right)^{2/p}\; ,\label{tt}
\ee
thereby generating the transition between the two Kaluza-Klein regions.
Furthermore, in accordance with ref.~\cite{letter,paper}, we find two
different types of solutions which are realized for the two different signs
of $\t$ in eq.~\refs{asympt}. The solutions with negative $t$ generally
end in a curvature singularity, whereas those with positive $t$ generally start
out from a curvature singularity. The two asymptotic Kaluza--Klein regions
and the transition period clearly exist for both types.

The interesting observation at this point is that the expansion powers and
Hubble parameters in the asymptotic regions are $SL(2)$ invariants. The
``r\^ole'' of $SL(2)$ is to determine the time of transition via
eq.~\refs{tt}. In addition, the $SL(2)$ parameters influence the detailed
behaviour during
the transition period. The main effect of applying $SL(2)$ to a given solution
is therefore to shift the transition time, that is, the time at which
the form is operative, according to eq.~\refs{tt}.

\vspace{0.4cm}

To illustrate the above discussion let us now concentrate on the subclass
of Friedman--Robertson--Walker (FRW)
cosmologies among our solutions. They are characterized by a 3--dimensional
isotropic spatial space with the Einstein frame metric
$\bar{G}_{mn}=a^2(\t )\d_{mn}$, where $a$ is the scale factor. From
eq.~\refs{solb4} and the fact that ${\rm det}(\cG )=1$, such a metric implies
that $\cG_{mn}=\d_{mn}$. Therefore, FRW cosmologies can be identified
as all solutions in the $SL(2)\times SL(3)$ equivalence classes with
$p_1=p_2=p_3=0$. Then the as yet unspecified parameter $p$ which
appears in the general solution~\refs{sol_M4} is in fact determined by
the constraints~\refs{p_cons4} to be
\be
 p = \sqrt{\frac{12}{7}}\; .
\ee
This means that there exists exactly one equivalence class with FRW
cosmologies. In particular, all FRW cosmologies are
$SL(2)\times SL(3)$--related and no FRW cosmology can be transformed into
a non--FRW cosmology (and vice versa) via $SL(2)\times SL(3)$. Using
eq.~\refs{solb4} we find for the scale factor
\be
 a = b_0^{7/2}(\g^2|\t |^{p+2}+\d^2|\t |^{-p+2})^{1/4}\; .
\ee
In the asymptotic Kaluza--Klein regions we can again express
the scale factor $a$ in terms of the comoving time $t$. For both asymptotic
regions we find for the Hubble parameter $H_a$
\be
 H_a\equiv\frac{1}{a}\frac{da}{dt}=\frac{1}{3t}\; .
\ee
This power law behaviour with power $1/3$ is typical for an evolution driven
by scalar field kinetic energy. Since the power is positive the universe
expands in the positive time branch $t>0$ and contracts in the
negative branch $t<0$. In both branches the form connects asymptotic
regions with the same expansion (or contraction) rate. As explained
before an $SL(2)$ transformation shifts the time of transition. Clearly,
$SL(2)$ would play an even more interesting r\^ole if the two
asymptotic regions had different properties, for example, if
one was expanding while the other was contracting. Though this is generically
the case for the general solution~\refs{Hub4}, requiring FRW--type solutions
forces us into a situation with equal evolution rates. This will be
different for the $D=5$ solutions of FRW type as discussed in the next
section.

\section{The Example $D=5$}

We are now dealing with a $\d =6$--dimensional internal manifold which
can be a torus $T^6$ or a Calabi--Yau 3--fold. The equations of motion
for the Lagrangian~\refs{L5} are given by the general
expressions~\refs{eoms}, \refs{Hrho} with $\d =6$ and $M=\cM$ inserted.
Here $\cM$ is the $SL(5)/SO(5)$ coset explicitly given in terms of the
metric and the 3--form in eq.~\refs{cos5}. From eq.~\refs{trafo} the
physical fields can be written as
\bea
 \bar{b} &=& \F^{-1/10}\, b \nn \\
 \bar{N} &=& \F^{-3/10}\, b^2 \label{trafo5}\\
 \bar{G}_{mn} &=& \F^{-2/5}\, b^4\, G_{mn}\nn\; .
\eea
The solution for the breathing mode can be read off from eq.~\refs{b_sol}
\be
 H=\frac{1}{6\t}\; ,\quad b=b_0\, |\t |^{1/6}\; .
\ee
For the coset $\cM$ we have from eq.~\refs{sol_M1} and \refs{p_cons}
\be
 \cM = P\, {\rm diag}(|\t |^{p_1},...,|\t |^{p_5})P^T\; ,\label{sol_M5}
\ee
with
\be
 \sum_{i=1}^5p_i=0\; ,\quad \sum_{i=1}^5p_i^2=\frac{10}{3}\label{p_cons5}
\ee
and $P\in SL(5)$. As before, we require $p_i\geq p_j$ for $i<j$. The
solution~\refs{sol_M5} contains 27 integration
constants. Three of them are given by the parameters $\{ p_i\}$ subject
to the constraints~\refs{p_cons5}, labeling the $SL(5)$ equivalence classes.
The remaining 24 integration constants parameterize the $SL(5)$ matrix
$P$ in eq.~\refs{sol_M5}.

What is the general physical picture emerging from the solution~\refs{sol_M5}?
The entries of the Einstein frame metric $\bar{G}_{mn}$ are given by a
sum of terms, each being of the form $K|\t |^q$. Here $q$ is
some power determined by the parameters $\{ p_i\}$ in~\refs{sol_M5} and $K$
is a combination of $SL(5)$ group parameters. As in the $D=4$ case, one expects
one of these terms to dominate over the others in certain ranges of $|\t |$.
For example, for a sufficiently large $|\t |$ the dominating term will
be the one with the maximal power $q$. In these asymptotic regions, the
evolution rates are controlled by the $\{ p_i\}$ and are therefore
$SL(5)$ invariant. As before, the $SL(5)$ parameters determine the
precise time range for the asymptotic regions and the time of transition.

Though generically clear, this is rather complicate to analyze in detail
for the general solution~\refs{sol_M5}. Therefore we concentrate
on the physically interesting case of FRW universes in the following. 
Consequently, we require a 3--dimensional spatial subspace of our
5--dimensional space to be isotropic. For the metric and the form field
this implies
\be
 G=\left(\ba{cc} c{\bf 1}_3&0\\0&\f\ea\right)\; ,\quad 
 {\bf B} = \left(\ba{c}{\bf 0}_3\\B\ea\right)\; ,\label{frw5}
\ee
where $c,\f ,B$ are time dependent scalars. Inserting~\refs{frw5} into
the coset parameterization~\refs{cos5} for $\cM$ results in
\be
 \cM =\left(\ba{cc}\cM_3&0\\0&\cM_2\ea\right)\; ,\quad
 \cM_3 = c\F^{-2/5}{\bf 1}_3\; ,\quad \cM_2 = \F^{-2/5}\left(\ba{cc}
         \f&-\f B\\ -\f B&\F +B^2\f\ea\right)\; ,\label{cos_frw}
\ee
with $\F =c^3\f$. Eq.~\refs{cos_frw} shows that, unlike in the
case $D=4$, the property ``FRW universe'' is not invariant under the full
U--duality group $SL(5)$. In particular, FRW universes can be mapped into
anisotropic solutions and vice versa using appropriate $SL(5)$
transformations. Since we wish to stay within the class of FRW solutions,
we should restrict ourselves to the subgroup
$H\equiv SL(3)\times SL(2)\times U(1)\subset SL(5)$ which leaves the structure
of $\cM$ in eq.~\refs{cos_frw} invariant. Explicitly, this subgroup acts
as
\be
 \cM_{2,3}\rightarrow P_{2,3}\cM_{2,3}P_{2,3}^T\label{Hact}
\ee
with $P_{2}\in GL(2)$, $P_{3}\in GL(3)$ and ${\rm det}(P_2){\rm det}(P_3)=1$.
Let us consider equivalence classes of solutions with respect to this
subgroup $H$ instead of the full group $SL(5)$. Then FRW universes are
specified by
\be
 p\equiv p_1=p_2=p_3
\ee
in eq.~\refs{sol_M5}. From eq.~\refs{p_cons5} we derive
\be
 p_{4,5} = -\frac{3p}{2}\pm\sqrt{\frac{5}{3}-\frac{15}{4}p^2}\; ,\quad
 |p|\leq \frac{2}{3}\; .
 \label{p45}
\ee
We have therefore found a one parameter set (with parameter $p$) of
$H$--inequivalent classes of FRW universes, each equivalence class for a
fixed $p$ spanned by the action of the group $H$ in eq.~\refs{Hact}.
How does $H$ act explicitly? First of all, $GL(3)\subset H$ is again
part of the global coordinate transformations and therefore trivial.
We concentrate on the $SL(2)$ part and write
\be
 P_2 = \left(\ba{cc} \a&\b\\ \g&\d\ea\right)\; ,\quad \a\d -\b\g = 1\; .
\ee
Then, $\cM_2$, $\cM_3$ take the form
\be
 \cM_2 = \left(\ba{cc}\a^2|\t |^{p_4}+\b^2|\t |^{p_5}&
                      \a\g |\t |^{p_4}+\b\d |\t |^{p_5}\\
                      \a\g |\t |^{p_4}+\b\d |\t |^{p_5}&
                      \g^2|\t |^{p_4}+\d^2|\t |^{p_5}\ea\right)\; ,
 \quad\cM_3 = |\t |^p{\bf 1}_3\; ,
\ee
where $p_{4,5}$ are given by~\refs{p45} in terms of the free parameter $p$.
By comparison with eq.~\refs{cos_frw} we can read off the expressions
for $c,\F ,B, \f$ and convert them to the physical fields via
eq.~\refs{trafo5}. The result is
\bea
 \bar{b} &=& b_0\, (\a^2|\t |^{p_4}+\b^2|\t |^{p_5})^{1/6}|\t |^{p/2+1/6}
            \nn \\
 \bar{N} &=& b_0^2\, (\a^2|\t |^{p_4}+\b^2|\t |^{p_5})^{1/2}|\t |^{3p/2+1/3}
  \nn \\
 \bar{G}_{mn} &=&a^2\,\d_{mn}\; ,\quad a = b_0^2\, (\a^2|\t |^{p_4}
                 +\b^2|\t |^{p_5})^{1/3}|\t |^{p+1/3} \\
 B &=& -\frac{\a\g |\t |^{p_4}+\b\d |\t |^{p_5}}
       {\a^2|\t |^{p_4}+\b^2|\t |^{p_5}}\; .\nn
\eea
As for $D=4$, we can solve for the comoving time $t$ in two asymptotic
regions leading to
\bea
 t &=& \frac{2b_0^2\, |\a |}{3p+p_4+8/3}|\t |^{3p/2+p_4/2+4/3}\, {\rm sgn}(\t )
       \; , \quad {\rm for}\quad |\t |\gg\t_{\rm form} \nn \\
 t &=& b_0^2\, |\b |\frac{2}{3p+p_5+8/3}|\t |^{3p/2+p_5/2+4/3}\, {\rm sgn}(\t )
       \; , \quad {\rm for}\quad |\t |\ll\t_{\rm form}\; ,
\eea
where
\be
 \t_{\rm form} = \left(\frac{\b}{\a}\right)^{\frac{2}{p_4-p_5}}\; .
\ee
In these regions, we find for the Hubble parameter $H_a$
\be
 H_a = \frac{P(p)}{t}\; ,\quad P(p) = \left\{\ba{lllll}
       \frac{2}{3}\frac{3p+p_4+1}{3p+p_4+8/3}&&{\rm for}
       &&|\t |\gg\t_{\rm form}\\
       \frac{2}{3}\frac{3p+p_5+1}{3p+p_5+8/3}&&{\rm for}
       &&|\t |\ll\t_{\rm form}\ea\right.\; .
       \quad \label{Hub5}
\ee
The expansion coefficient $P(p)$ depends on the free parameter $p$ and is
generically different in the two asymptotic regions. As can be seen from
fig.~1, it is always positive for large $|\t |\gg \t_{\rm form}$
and can have both signs for $|\t |\ll \t_{\rm form}$. For the
positive branch $t>0$ this implies a universe which is expanding or
contracting at early time and is turned into an expanding universe at late
time. The situation for the negative branch is reversed; the universe
is always contracting at early time ($|t|$ large and $t<0$) and can
be contracting or expanding later. As before, the Hubble parameter~\refs{Hub5}
and hence the aforementioned properties are $SL(2)$ invariant. The transition
time given by
\be
 |\t |\sim\t_{\rm form}=\left(\frac{\b}{\a}\right)^{\frac{2}{p_4-p_5}}\; ,
\ee
on the other hand, depends on $SL(2)$ parameters along with the details of
the transition.\\
[-0.5cm]

\epsfbox[-80 0 500 210]{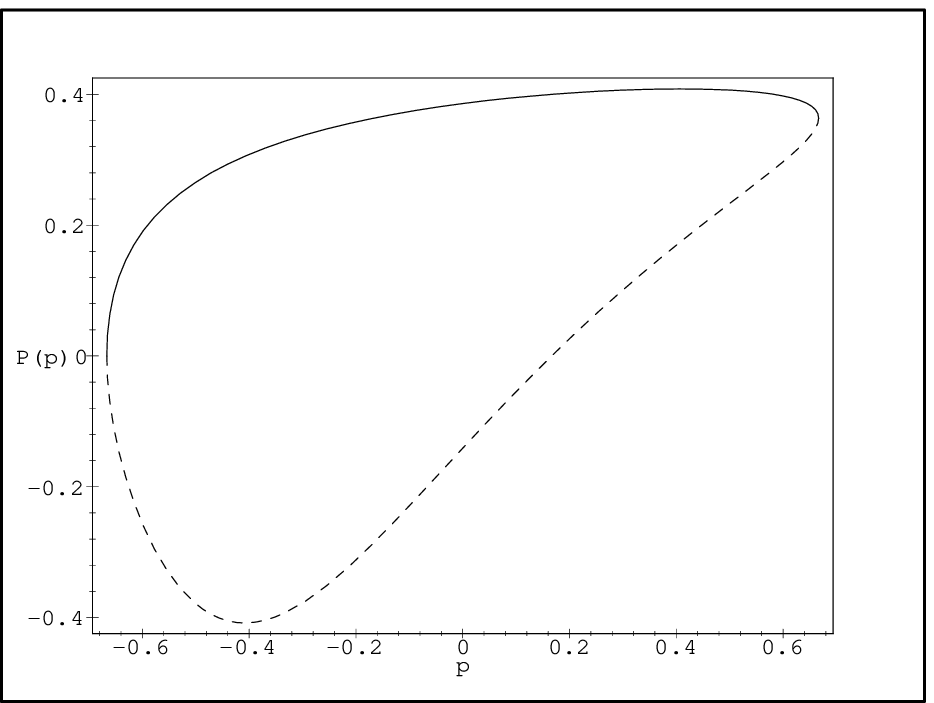}
\centerline{\em Fig 1: Expansion coefficient $P(p)$ for
$|\t |\gg\t_{\rm form}$ (solid curve) and $|\t |\ll\t_{\rm form}$
(dashed curve).}
\vskip 0.4cm

Suppose now, we choose a solution in the positive branch which is contracting
for a short period of time and then turns into expansion. By applying
appropriate $SL(2)$ transformations to this solution the contraction period
can be made arbitrarily long. An additional time reversal $t\rightarrow -t$
leads to a negative branch solution with an expansion period that can be
arbitrarily long. The extreme limits are possible. By choosing $P_2={\bf 1}$
($\b =0$ in particular) we have a positive branch solution which is always
expanding. As the other extreme we may set
\[
 P_2 = \left(\ba{cc}0&1\\-1&0\ea\right)
\]
($\a =0$ in particular) which
generates an expanding negative time branch solution. We have therefore shown
that a combination of U--duality and time reversal can map expanding
negative and positive time branch solutions into each other. The expansion
coefficients in the negative and positive branch then correspond to the
lower and upper part of the curve in fig.~1. An analog mapping, carried
out by a T--duality transformation combined with a time reversal, is the
starting point of the pre--big--bang scenario~\cite{pbb} of weakly coupled
heterotic string cosmology.

\section{Conclusion}

In this paper, we have reduced 11--dimensional supergravity to $D$
dimensions keeping a ``minimal'' field content in the low energy theory,
that is, the $D$--dimensional metric and 3--form and the breathing mode
of the internal Ricci--flat manifold. We have shown that cosmological
solutions of such a $D$--dimensional effective theory are transformed into
each other by the ``external'' subgroup of U--duality which is identical
with the U--duality group of $12-D$--dimensional maximal supergravity.
Based on this fact, we have worked out a U--duality invariant 1--dimensional
effective action for cosmological models and determined its general
solution written in a U--duality covariant way. These results have been
explicitly applied to the cases $D=4,5$. For these examples we recover
some of the results obtained previously. Generally, we find  positive and
negative time branch solutions. The form field in each branch is operative
during a finite period of time only, thereby generating a transition
between two asymptotic Kaluza--Klein regions.

\vspace{0.4cm}

A new property in connection with the U--duality covariant approach which
we found is the invariance of certain characteristic cosmological
properties such as Hubble parameters and expansion powers in the asymptotic
Kaluza--Klein regions. The r\^ole of U--duality is to fix the time and
the details of the transition generated by the 3--form. A U--duality
transformation, acting on a given solution, therefore shifts the time
of transition between the Kaluza--Klein regions but does not affect the
evolution in these regions.

One is tempted to interpret such a phenomenon as asymptotic U--duality
invariance of the cosmological entropy, as measured by the area of the
cosmological horizon. There is a problem, however, with such an
interpretation. First of all, the cosmological horizons, defined by
\be
 \l_{\rm part}(t) = a(t)\int_{t_i}^t\frac{dt'}{a(t')}\; ,\quad
 \l_{\rm event}(t) = a(t)\int_{t}^{t_f}\frac{dt'}{a(t')} \label{hor}
\ee
are nonlocal in time. Here $\l_{\rm part}$, $\l_{\rm event}$ are the particle
and the event horizon, respectively. The times $t_i$, $t_f$ are the maximal
past and future extensions of the solution. For our solutions we always
have $t_i=0$, $t_f=\infty$ for the positive branch and $t_i=-\infty$, $t_f=0$
for the negative branch. Moreover, the typical behaviour of the scale factor
$a(t)$ in the asymptotic regions is given by a power law $a(t)\sim t^q$, where
$|q|<1$. Therefore, the only finite horizons are the particle horizon for
the positive branch and the event horizon for the negative branch.
Consider, for example, the particle horizon $\l_{\rm part}(t)$ in the
positive branch at a given time $t$ and an equivalence class of U--duality
related solutions.
The solutions in this class will have different times $t_{\rm form}$ at
which the form is operative and there will be solutions with $t<t_{\rm form}$
as well as $t>t_{\rm form}$ in this class. For the latter, the integral in
eq.~\refs{hor} extends over the transition period and therefore depends
on U--duality group parameters. As a result, though the Hubble
parameter $H_a\equiv\frac{1}{a}\frac{da}{dt}$ is U--duality invariant, this
is not the case for the particle horizon as defined in eq.~\refs{hor}.
Despite this problem with a thermodynamical interpretation,
we find the asymptotic U--duality invariance of relevant cosmological
quantities remarkable.

Another important result in $D=5$ is, that a combination of U--duality and
time reversal can be used to map expanding negative and positive time
branch solutions into each other. Therefore, a scenario analogous to
T--duality pre--big--bang cosmology appears to be a possibility within
our models.

\vspace{0.4cm}

{\bf Acknowledgments} A.~L.~is supported by a fellowship from Deutsche
Forschungsgemeinschaft (DFG). A.~L.~and B.~A.~O.~are supported in part by
DOE under contract No.~DE-AC02-76-ER-03071.
\end{document}